\documentclass[12pt]{iopart}
\usepackage{graphicx}
\usepackage{cite}
\usepackage{amsthm}
\usepackage{iopams}
\usepackage{hyperref}
\begin{document}
\title[A simple model for the spin-singlet]{A simple model for the spin-singlet: mathematical equivalence of non-locality, slave will, and conspiracy}
\author{Antonio {Di Lorenzo}}
\address{Instituto de F\'{\i}sica, Universidade Federal de Uberl\^{a}ndia, 38400-902 Uberl\^{a}ndia, Minas Gerais, Brazil}
\begin{abstract}
A  hidden variable model reproducing the quantum mechanical probabilities for a spin singlet is presented. 
The model violates only the hypothesis of independence of the distribution for the hidden variables from the detectors settings and \emph{vice versa} (Measurement Independence). 
It otherwise satisfies the hypotheses of Setting Independence, Outcome Independence -- made in the derivation of Bell inequality -- and that of 
compliance with Malus's Law -- made in the derivation of Leggett inequality. 
It is shown that the violation of the Measurement Independence hypothesis may be explained alternatively by assuming a non-local influence of the detectors settings on the hidden variables, or by taking the hidden variable to influence the choice of settings (limitation of free will), or finally by purporting a conspiracy. 
It is demonstrated that the last two cases admit a realization through existing local classical resources. 
\end{abstract}
\maketitle
\section{Introduction.} 
Hidden-variable model, or theory, is the (unfortunate) name with which physicists refer to 
a hypothetical theory where the quantum state $\Psi$ (or $\rho$ for mixtures) of a system 
is supplemented by additional parameters, $\lambda$. 
Quantum mechanics provides a set of rules that determine the probabilities of observing 
events $e$ for a given preparation of the system $\Psi$ and a given experimental setup $\Sigma$, 
$P^{QM}(e|\Psi,\Sigma)$. 
The challenge for the hidden-variable theory is to find a 
distribution of the $\lambda$, $\mu(\lambda|\Psi,\Sigma)\ge 0$, and a set of conditional probabilities 
$P(e|\lambda,\Psi,\Sigma)$ such that the quantum mechanical predictions are reproduced on average, namely
$P^{HV}(e|\Psi,\Sigma)=P^{QM}(e|\Psi,\Sigma)$, 
where  $P^{HV}(e|\Psi,\Sigma)\equiv\int\!d\lambda\, \mu(\lambda|\Psi,\Sigma) P(e|\lambda,\Psi,\Sigma)$, as follows from Bayes\rq{}s rule \cite{Bayes1763}.  
Alternatively, considering that experiments have some unavoidable imprecision, one could require just that 
$P^{HV}(e|\Psi,\Sigma)\simeq P^{QM}(e|\Psi,\Sigma)$ so that the the hidden-variable model 
 reproduces the experimental data with an accuracy comparable to that of quantum mechanics. 
The earliest example of a hidden-variable theory is given by the Bohm formulation \cite{Bohm1952a,Bohm1952b}. 

As demonstrated by Bell \cite{Bell1964}, 
all hidden variable theories satisfying three hypotheses --- known as Measurement Independence (Uncorrelated Choice in our terminology to be introduced below), 
Setting Independence, and Outcome Independence (Reducibility of Correlations, in our terminology) --- are incompatible with quantum mechanics and, more importantly, 
with experimental evidence \cite{Rowe2001}. 
More recently, Leggett \cite{Leggett2003} demonstrated the incompatibility of quantum mechanics with all theories satisfying Measurement Independence and 
Malus\rq{}s law. These theories as well were ruled out by experiment \cite{Lee2011}. 
By violating one or more hypotheses, however, it is possible to reproduce the quantum mechanical predictions. 
Examples that the violation of Measurement Independence can lead to models reproducing the quantum mechanical prediction 
were provided in Refs.~\cite{Shimony1985a,Brans1988,Hall2010}. 
The amount of violation of Measurement Independence necessary to reproduce quantum mechanics 
was recently quantified \cite{Hall2010,Barrett2011}. 
In the present paper, we provide a model that satisfies at the same time 
the hypotheses Setting Independence, Outcome Independence, and Malus\rq{}s Law, but not Measurement Independence. 
For comparison, in the literature there are models violating Measurement and Outcome Independence \cite{Cerf2005,Groblacher2007b}, 
Measurement Independence and Malus\rq{}s Law \cite{Brans1988,Hall2010}, Setting Independence \cite{Toner2003}, 
and the only model satisfying Setting Independence , Outcome Independence, and Malus\rq{}s Law \cite{DeZela2008} turns out to be flawed \cite{DiLorenzo2012c}.\footnote{The distribution $\mu(\lambda|\Sigma)$ is not normalized to one; if it was, then the correlator would be 
$-\mathbf{a}\cdot\mathbf{b}/4$, four times smaller than the quantum mechanical one.
} 

\section{Some definitions.}
In order to discuss the hypotheses underlying Bell and Leggett inequalities introduced below, we introduce some useful concepts. 
We distinguish two kinds of hidden variables: 
global parameters and local parameters. The former ones can not be ascribed to a region of spacetime, while the latter ones can. 
Furthermore, local parameters can be detected by a single-shot measurement, while global ones require an ensemble of measurements or 
the specification of a preparation procedure. 
In other words, the value of local parameters can constitute an event. 

For instance, consider a card chosen at random from a deck; the deck itself is chosen at random from a set of decks, each having a 
different distribution of red and black cards. The probability of the card being red depends on which deck was chosen. 
This information constitutes a global parameter, since it can not be ascribed to the card. On the other hand, the card possesses the property of being black or red before 
it is measured. This property is a local parameter. The wave-function of a quantum system is a global parameter. 
According to the naive, classical world-view, the knowledge of all the local parameters makes global parameters irrelevant. 
On the other hand, the shift in the epistemic paradigm introduced by quantum mechanics consists in recognizing that some global parameters 
cannot be simply reduced to the ignorance of some fundamental yet unknown local parameters, and that the events resulting 
as the outcome of a measurement are not interpretable as preexisting local parameters belonging to the observed system. 

Now we are in a position to define \emph{locality}, by which we mean the impossibility of action-at-a-distance (NAD). 
The very word \lq\lq{}action\rq\rq{} stems from classical determinism, and indicates the change in a local parameter. However, we need to extend the concept of locality 
to probabilistic theories. Our proposed definition is: locality implies that  
the probability of observing an event $e$ at a spacetime region $R$ can 
be expressed by a function that depends only 
on global parameters and on those local parameters 
that are localized within the region $R$. 
In formulas 
\begin{equation}\label{eq:locdef}
P(e_R|\Psi,\Sigma,\lambda_G,\lambda_A,\lambda_B,\cdots \lambda_Z) = P(e_R|\Psi,\Sigma_R,\lambda_G,\lambda_R). 
\end{equation}
Eq.~\eref{eq:locdef} should be interpreted as a procedure (in the algorithmic sense) assigning a value $P$ according to a routine that takes as 
an input the variables $\Psi,\Sigma_R,\lambda_G,\lambda_R$, which are not to be confused with their values. 
Clearly, if the local parameters are correlated, the value of $\lambda_R$ may coincide, 
for instance, with that of $\lambda_S$. However, a local operation in $S$ will change the value of $\lambda_S$ while keeping 
$\lambda_R$ fixed, and this will not affect the marginal probability at $R$. 
Finally, 
if the event consists in the measurement of the value of a  local parameter, say of $\lambda_R=L_R$, for which we assume there is 
a measuring device $\Lambda_R$, then 
\begin{equation}
P(L_R|\Psi,\Sigma,\lambda_G,\Lambda_A,\Lambda_B,\cdots \Lambda_Z) = P(L_R|\Psi,\lambda_G,\Lambda_R). 
\end{equation}
The parameters and the settings are all considered to be calculated at the same time $t$. 
No hypothesis is made about the equations of motion for the additional parameters. 
\section{The setup.}
\begin{center}
\begin{figure}
\includegraphics[width=5in]{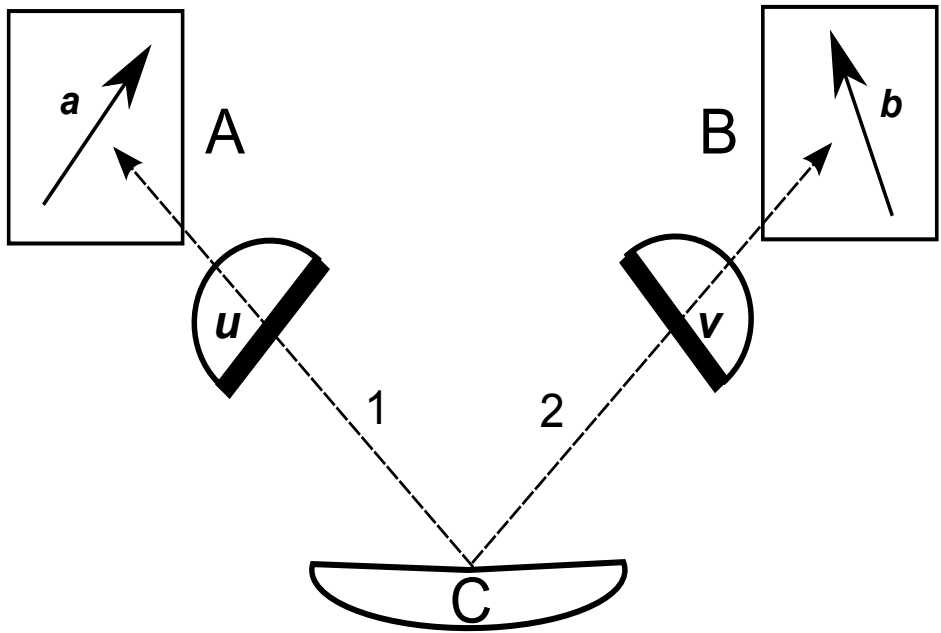}
\caption{
\label{fig:eprbsetup} Scheme of the setup considered.Regions A and B are spacelike separated. The semicircles represent hypothetical detectors for the hidden variables as discussed in the text.}
\end{figure}
\end{center}
The setup considered (see Fig. \ref{fig:eprbsetup}) consists of two particles produced in a region $C$ of the 
spacetime, each travelling to a different detection region, $A$ and $B$. The measurements of the two particles does 
not need to happen at the same time, but the two detection events are assumed to have spacelike separation, 
so that a reference frame exists in which the measurements are simultaneous. 
The outcomes of the two measurements are two-valued, and will be denoted by 
$e_A=\sigma=\pm 1$ and $e_B=\tau=\pm 1$. 
The $A$ ($B$) detector is characterized by a unit vector $\mathbf{n}_A=\mathbf{a}$ ($\mathbf{n}_B=\mathbf{b}$), corresponding to the 
orientation of a spin measuring device in Quantum Mechanics. 
The quantity of interest is the joint probability $P(\sigma,\tau|\Psi,\Sigma)$
with $\Psi$ describing the preparation of a singlet state, and $\Sigma=\{\mathbf{a},\mathbf{b}\}$ 
specifying the observables being measured. 
In the following, we shall write the hidden variables as $\lambda=(\lambda_G,\lambda_A,\lambda_B)$, 
where $\lambda_G$ refers to the global parameters, and $\lambda_j$ ($j\in\{A,B\})$ to the local parameters.
Since $\Psi$ appears as a prior in all the probabilities considered, it is omitted for brevity. 
\section{Overview of previous results}
\subsection{Hypotheses at the basis of Bell inequality.}
References~\cite{Bell1964,Clauser1969,Bell1971,Clauser1974} discriminated between quantum mechanics 
and some special classes of hidden-variable theories through inequalities, known as Bell, CHSH, and CH inequalities, 
which have been verified experimentally \cite{Rowe2001}. 
There are three hypotheses needed to derive Bell-type inequalities: 
\begin{enumerate}
\item
Measurement Independence, i.e., the marginal distribution of the local parameters 
and the choice of the corresponding remote settings are uncorrelated,  
$\mu(\lambda|\Sigma)=\mu(\lambda)$, with $\Sigma=\{\mathbf{a},\mathbf{b}\}$ 
representing the detectors settings. 
Herein, I shall refer to this hypothesis with the more descriptive term \lq\lq{}Uncorrelated Choice\rq\rq{}. 
\item
Setting Independence, i.e. the marginal probability of observing an event at station $j$ does not depend on the setting of the remote station, 
$P_A(\sigma|\lambda,\Sigma)=P_A(\sigma|\lambda,\mathbf{a})$. 
\item 
Outcome Independence, i.e., for fixed $\lambda$ the events $\sigma$ and $\tau$ are uncorrelated,  
\begin{equation}\label{eq:rc}
Q_B(\tau|\sigma,\lambda,\Sigma)=P_B(\tau|\lambda,\Sigma) ,
\end{equation}
where the marginal and conditional probabilities are by definition 
\begin{eqnarray}
\label{eq:marg}
P_B(\tau|\lambda,\Sigma)
&\equiv  
 \sum_{\sigma\rq{}} P(\sigma',\tau|\lambda,\Sigma) ,
\\
\label{eq:cond}
Q_B(\tau|\sigma,\lambda,\Sigma)
&\equiv
 \frac{P(\sigma,\tau|\lambda,\Sigma)}{\sum_{\tau\rq{}}P(\sigma,\tau\rq{}|\lambda,\Psi)} .
\end{eqnarray}
Reference~\cite{Shimony1990} coined the name ``Outcome Independence'' for the additional hypothesis \eref{eq:rc}, 
while Jarrett \cite{Jarrett1984} and Bell \cite{Bell1990} referred to it as \lq\lq{}completeness\rq\rq{}. 
In my opinion, the latter term is too ambiguous, while the former is too technical. I shall hence refer to this hypothesis as \lq\lq{}Reducibility of Correlations\rq\rq{}, 
since it amounts to assume that for given $\lambda$ there are no correlations between the two outcomes. 
\end{enumerate}

We notice that Uncorrelated Choice/Measurement Independence is stronger than locality, 
which implies more generally $\mu_A(\lambda_A|\Sigma)\equiv \int d\lambda_G d\lambda_B \mu(\lambda_G,\lambda_A,\lambda_B|\Sigma)=\mu_A(\lambda_A|\mathbf{a})$, 
and Setting Independence weaker, since it allows a dependence of the marginal 
probability on the remote local setting $\lambda_B$. Thus, as $\lambda_B$ could be changed at station $B$, and 
this would change the marginal probability at station $A$,  
Setting Independence would allow instantaneous communication from $B$ to $A$. 
Hence, Setting Independence does not imply no-signaling, contrary to a widespread belief. 

Concerning the hypothesis Reducibility of Correlations/Outcome Independence, it has been demonstrated \cite{Hall2011} that any model satisfying it can be supplemented by further additional parameters $\lambda\rq{}$ in such a way 
that the new model becomes deterministic, i.e. $P(e_j|\{\lambda,\lambda\rq{}\},\Sigma)\in\{0,1\}$. 
However, experiment must provide the ultimate test for any theory, i.e., if one formulates a model and in addition 
gives a prescription to either fix or measure $\lambda$, then 
the probability $P(e_j|\lambda,\Sigma)$ becomes experimentally accessible and the theory falsifiable. It may happen 
that there is no way to measure or fix 
the additional parameters $\lambda\rq{}$, so that the deterministic completion would turn out to be a useful fiction.

\subsection{Hypotheses at the basis of Leggett inequality.}
Reference~\cite{Leggett2003} considered a class of hidden-variable models that do not necessarily satisfy
Reducibility of Correlations/Outcome Independence, but obey an analogue of Malus's Law for the hidden variables, 
$P(e_j|\lambda,\Sigma)=(1+e_j\mathbf{u}_j\cdot\mathbf{n}_j)/2$, where $\lambda=[\mathbf{u}_A,\mathbf{u}_B]$ 
consists in two unit vectors, localized at particle $A$ and $B$, 
and $\mathbf{n}_j$ is a unit vector denoting the setting of station $j$. 
Thus Malus\rq{}s Law is a special case of Setting Independence. 
It was shown \cite{Leggett2003,Groblacher2007b,Branciard2008} that these models 
predict a correlator satisfying an inequality known as Leggett inequality, which is violated by quantum mechanics and 
by experiments \cite{Lee2011}. 
The assumption of Malus\rq{}s Law appeals to the intuitive notion that each spin possesses a vector $\mathbf{u}$ describing its polarization and influencing the outcome of the measurement according 
to the well known Malus\rq{}s law, which applies for pure single-particle states. 

In summary, Bell inequalities are obtained assuming that Uncorrelated Choice/Measurement Independence, Setting Independence, and Reducibility of Correlations/Outcome Independence hold, 
while Leggett inequalities are obtained assuming Uncorrelated Choice and compliance with Malus\rq{}s Law.

\section{Model and results.} 
By violating one of the hypotheses at the basis of Bell and Leggett inequalities it may 
be possible to violate them. Reproducing quantum mechanics, however, is not guaranteed, 
since the violation of Bell and Leggett inequalities is a necessary 
but not sufficient condition. 
Here we provide a model that not only violates the inequalities, but also reproduces the quantum mechanical prediction for a spin singlet. 
Another distinguishing feature of the model discussed herein is the 
simplicity of the distribution $\mu$ [compare Eq.~\eref{eq:hiddvardens} below], which is not 
contrived \emph{ad hoc} in order to reproduce the quantum mechanics of a spin-singlet. 
Let us consider the following hidden-variable model: 
The hidden variables consist of two unit vectors $\lambda=\{\mathbf{u},\mathbf{v}\}$, 
the first being associated with the particle going to $A$ 
and the second with the particle going to $B$. 
The joint probability, conditioned on the values $\mathbf{u},\mathbf{v}$ is 
\begin{equation}
P(\sigma,\tau|\mathbf{u},\mathbf{v},\Sigma)=
\frac{1}{4}\left(1+\sigma \mathbf{u}\cdot\mathbf{a}\right)
\left(1+\tau \mathbf{v}\cdot\mathbf{b}\right) \ ,
\end{equation}
so that the marginal and conditional probabilities are
\begin{eqnarray}
\label{eq:margprob}
&P(\sigma|\mathbf{u},\mathbf{v},{\Sigma})=
\frac{1}{2}\left(1+\sigma \mathbf{u}\cdot\mathbf{a}\right) \ , \\
\label{eq:condprob}
&P(\tau|\sigma,\mathbf{u},\mathbf{v},{\Sigma})=
\frac{1}{2}\left(1+\tau \mathbf{v}\cdot\mathbf{b}\right) \ .
\end{eqnarray}
The model obeys Setting Independence, since the probability of finding outcome $\sigma$ 
depends solely on the variable $\mathbf{u}$ associated to the particle at $A$, and in this sense the marginal probability obeys the locality condition. 
Furthermore, Eq.~\eref{eq:margprob} 
states that the Malus\rq{}s law is obeyed, while Eq.~\eref{eq:condprob}
shows that Reducibility of Correlations/Outcome Independence is satisfied as well. 
The hidden variables have the following probability density
\begin{eqnarray}
\label{eq:hiddvardens}
&\mu(\mathbf{u},\mathbf{v}|{\Sigma})=
\frac{1}{4} \sum_{\mathbf{p}=\pm \mathbf{a},\pm \mathbf{b}}
 \delta(\mathbf{u}-\mathbf{p}) \delta(\mathbf{v}+\mathbf{p})
\ .
\end{eqnarray}
It is immediate to verify that upon integration over the hidden variables
\begin{eqnarray}\nonumber
P^{HV}(\sigma,\tau|{\Sigma})\equiv& 
\int d\mathbf{u}d\mathbf{v} \mu(\mathbf{u},\mathbf{v}|{\Sigma})P(\sigma,\tau|\mathbf{u},\mathbf{v},\Sigma)\\
=&\frac{1}{4}\left(1-\sigma\tau \mathbf{a}\cdot\mathbf{b}\right),
\end{eqnarray}
coincides with the quantum mechanical predictions for a spin singlet. 
The reason that, in particular, Bell and Leggett inequalities are violated by the model above is apparent: 
the distribution of the hidden variables and that of the settings of the detectors are correlated i.e. (UC) does not hold in our model.

\subsection{Non-local realization of the model.} 
Let us assume that $\mathbf{u}$ and $\mathbf{v}$ are local parameters. Then Eq.~\eref{eq:hiddvardens} 
violates the principle of locality, 
since, e.g., the marginal probability for $\mathbf{u}$ is
\begin{equation}
\mu_A(\mathbf{u}|\Sigma)=\frac{1}{4}\sum_{\alpha=\pm 1}\left[\delta(\mathbf{u}+\alpha\mathbf{a})+\delta(\mathbf{u}+\alpha\mathbf{b})\right] . 
\end{equation}
We recall that $\mathbf{u}$ and the settings $\mathbf{a},\mathbf{b}$ are evaluated at the same time. 
Thus, a change in $\mathbf{b}$ can influence instantaneously the distribution of the remote parameter $\mathbf{a}$, 
and hence the non-locality. 
We prove that in this case there can be instantaneous communication between the regions $A$ and $B$, 
provided that the hidden parameters are measurable. 
Suppose that two observers at $A$ 
and $B$ agree on two orientations $\mathbf{a}$, $\mathbf{b}$, e.g., orthogonal to each other. They use the following protocol: 
immediately before the particles impinge on the spin detectors, they measure the hidden variables 
(see Fig. \ref{fig:eprbsetup}); 
if the observer $A$ measures $\mathbf{u}=\mathbf{a}$, 
(i.e., the orientation is determined by the one of the spin detector in $L$), 
she will turn her apparatus in the direction $+\mathbf{b}$ if she wants to make sure that the observer in $B$ obtains 
the result $-1$, which is agreed to correspond to the 0-bit, or in the direction $-\mathbf{b}$ 
if she wants to make sure that the observer in $B$ obtains 
the result $+1$, which is agreed to correspond to the 1-bit; 
if instead the hidden parameters turns out to be $\mathbf{u}=-\mathbf{a}$, 
the observer in $A$ will make the opposite switching; the observer at $B$, on the other hand, will take no action 
whenever he measures $\mathbf{v}=\pm\mathbf{a}$, since he knows he will be on the receiving end of the transmission; 
when instead $\mathbf{v}=\pm\mathbf{b}$ the observer in $B$ will switch 
his apparatus in an analogous fashion as $A$ does in the other cases,  
so that he can send instantaneous information to $A$, who will be measuring $\mathbf{u}=\mp\mathbf{b}$ and knows that she should take no action in this case.  

\subsection{Local realization.}
We remark that the non-locality of the model does not follow automatically from the hypothesis in 
Eq.~\eref{eq:hiddvardens}, but stems from the assumption that the settings of the detectors determine $\mathbf{u},\mathbf{v}$ and not \emph{vice versa}. Since probability theory is time-symmetric and acausal, one could assume the opposite cause-effect relation as 
in Refs. \cite{Brans1988,Hall2010}. 
The model could then be 
reproduced through classical resources in the 
following way: The detector at $A$ ($B$) and the entangler $C$ share a 
pseudo-random \footnote{In classical mechanics it is impossible to have genuine shared randomness between distant parties.} number generator, 
which gives as output a unit vector $\mathbf{m}$ ($\mathbf{n}$). The random-number generator (RNG) at $A$ and $B$ 
is delayed by a time $\tau$ equal to the time-of-flight of the particles, respect to the one at $C$.  
With probability $1/4$ the entangler produces one of four possible pairs $(\pm\mathbf{m},\mp\mathbf{m}), (\pm\mathbf{n},\mp\mathbf{n})$, then attaches 
the first member of the pair to the particle reaching $A$, and the second to the one reaching $B$, thus forcing $\mathbf{u}=-\mathbf{v}\in\{\pm \mathbf{m},\pm\mathbf{n}\}$. 
At the moment of choosing the direction along which to measure, $A$ and $B$ consult their RNG and use its outcome, setting 
$\mathbf{a}=\mathbf{m}$ and $\mathbf{b}=\mathbf{n}$. 
The outcome of each measurement is then randomized independently as for Eq.~\eref{eq:margprob}. 
In formulas 
\begin{eqnarray}
\nonumber
&\mu(\mathbf{m},\mathbf{n},\alpha,\beta,\mathbf{u},\mathbf{v}|{\Sigma})=
\delta(\mathbf{u}+\mathbf{v})
\delta(\mathbf{m}-\mathbf{a})
\delta(\mathbf{n}-\mathbf{b})
\\
&\times
\frac{1}{4} \left[\frac{1-\alpha}{2} \delta(\mathbf{u}-\beta\mathbf{m})+ 
\frac{1+\alpha}{2} \delta(\mathbf{v}-\beta\mathbf{n})
\right]
\label{eq:hiddvardens2}
\ ,
\end{eqnarray}
where $\alpha,\beta=\pm 1$, $\mathbf{m},\mathbf{n}$ are global parameters, and $\mathbf{u},\mathbf{v}$ local parameters. 
The weak point of such an explanation
 is not the presumed violation of free will (all in all free will is limited by physical laws, and 
it could be but an illusion if these are deterministic), but the origin of the correlations
 and the persistence of detector-entangler correlations 
notwithstanding the effects of the environment. Indeed, at some point in the common past light-cone of $A$ (or $B$) and $C$, the two shared a random number generator. 
Yet, all events that are in the past light-cone of $A$ and can not be causally correlated with $C$ turn out to be irrelevant in the determination of $\mathbf{a}$, even 
though they are more recent than the sharing of the RNG.

Furthermore, if the detectors at $A$ and $B$ were two automata complex enough to possess self-awareness, 
but without the possibility of finding out their inner workings, and these automata could
 measure $\mathbf{u}$ and $\mathbf{v}$,  
they would not only believe that they were acting out of free will, but also that they could establish 
superluminal communication according to the protocol illustrated previously, 
while a wary external observer would see them reciting a predetermined script. 
This consideration sheds a new light on the issue of \lq\lq{}free will\rq\rq{}: If our choices were determined by underlying variables, which we were able to measure (i.e. if we were puppets 
who could see their strings), 
then we would be able to test whether we have free will by trying to send superluminal signals. In case the communication resulted to be botched, we would have evidence in favor of 
\lq\lq{}slave will\rq\rq{} otherwise we would observe instantaneous communication of meaningful information \footnote{We exclude the unfalsifiable hypothesis of 
a cosmic choreography, where the parameters would determine not only our immediate  actions, 
but also the association of meaning to some special sequences of bits (language).}.

Finally, the same probability distribution would arise if $A$ and $B$ were conscious agents aware of the variables $\mathbf{m},\mathbf{n}$ 
and choosing their settings accordingly, in order to produce the quantum correlations. Assuming that they could choose not to do so, we would have then 
not a violation of \lq\lq{}free will\rq\rq{}, but a conspiracy, leading to the same probability distribution. 
This shows that it is not possible to deduce, from the assumed violation of UC, which among the hypotheses
 \lq\lq{}free will\rq\rq{}, \lq\lq{}locality\rq\rq{}, or \lq\lq{}no-conspiracy\rq\rq{} is being violated, and conversely, that none 
of the three hypotheses implies by itself UC, unless additional assumptions about the physical nature of the hidden variables are made.

\section{Conclusions.} 
The hidden-variable model presented  does not only violate Bell and Leggett inequalities, but 
reproduces the results of Quantum Mechanics for a spin singlet. The model 
violates only the hypothesis of uncorrelated choice, but satisfies all other assumptions at the basis of Bell and Leggett 
inequalities, namely Setting Independence, Reducibility of Correlations/Outcome Independence, and compliance with Malus's Law. 
Thanks to this, the model seems to appeal to our intuitive, classical notion of spin polarization: both particles have 
a fixed value of the polarization, which determines the probability of each experimental outcome. 
It can be realized indifferently by an alleged non-local influence of the detector on the hidden variables, or 
by preexisting correlations between the entangler and the settings of the stations, which could be seen either 
as a conspiracy or as a limitation of free will. Thus mathematical hypotheses about the form of probabilities cannot be 
claimed to derive from physical requirements, unless the variables appearing in the model are given first a physical meaning. 

This work was supported by Funda\c{c}\~{a}o de Amparo \`{a} Pesquisa do 
Estado de Minas Gerais through Process No. APQ-02804-10.

\section*{References}
\providecommand{\newblock}{}

\end{document}